\documentclass[useAMS,usenatbib]{mn2e_mod}
\usepackage[utf8]{inputenc}
\usepackage{amsmath}
\usepackage{amsfonts}
\usepackage{amssymb}
\usepackage{graphicx}
\usepackage[a4paper]{geometry}
\usepackage{tablefootnote}
\usepackage{deluxetable}
\usepackage{natbib}
\newcommand{\Msol}{\ensuremath{\mathrm{M_{\odot}}}}

\title [Bright but slow]{Bright but slow -- Type II supernovae from OGLE-IV -- Implications for magnitude limited surveys}

		\author[Poznanski et al.]
		{D. Poznanski$^{1}$\thanks{dovi@tau.ac.il},
		 Z. Kostrzewa-Rutkowska,$^{2}$,
		 L. Wyrzykowski$^{2,3}$,
		\newauthor \& N. Blagorodnova$^{3}$\\
		\\
		$^{1}$School of Physics and Astronomy, Tel-Aviv University, Tel Aviv 69978, Israel.\\
		$^{2}$Warsaw University Observatory Al. Ujazdowskie 4, 00-478 Warszawa, Poland\\		
        $^{3}$Institute of Astronomy, University of Cambridge, Madingley Road, Cambridge CB3 0HA, UK\\
		}

\begin{document}
	\maketitle
	\label{firstpage}
	\begin{abstract}
We study a sample of 11 Type II supernovae (SNe) discovered by the OGLE-IV survey. All objects have well sampled $I$-band light curves, and at least one spectrum. We find that 2 or 3 of the 11 SNe have a declining light curve, and spectra consistent with other SNe II-L, while the rest have plateaus that can be as short as 70\,d, unlike the 100\,d typically found in nearby galaxies. The OGLE SNe are also brighter, and show that magnitude limited surveys find SNe that are different than usually found in nearby galaxies. We discuss this sample in the context of understanding Type II SNe as a class and their suggested use as standard candles. 

\end{abstract}

\begin{figure*}
\centering
\includegraphics[width=.95\textwidth]{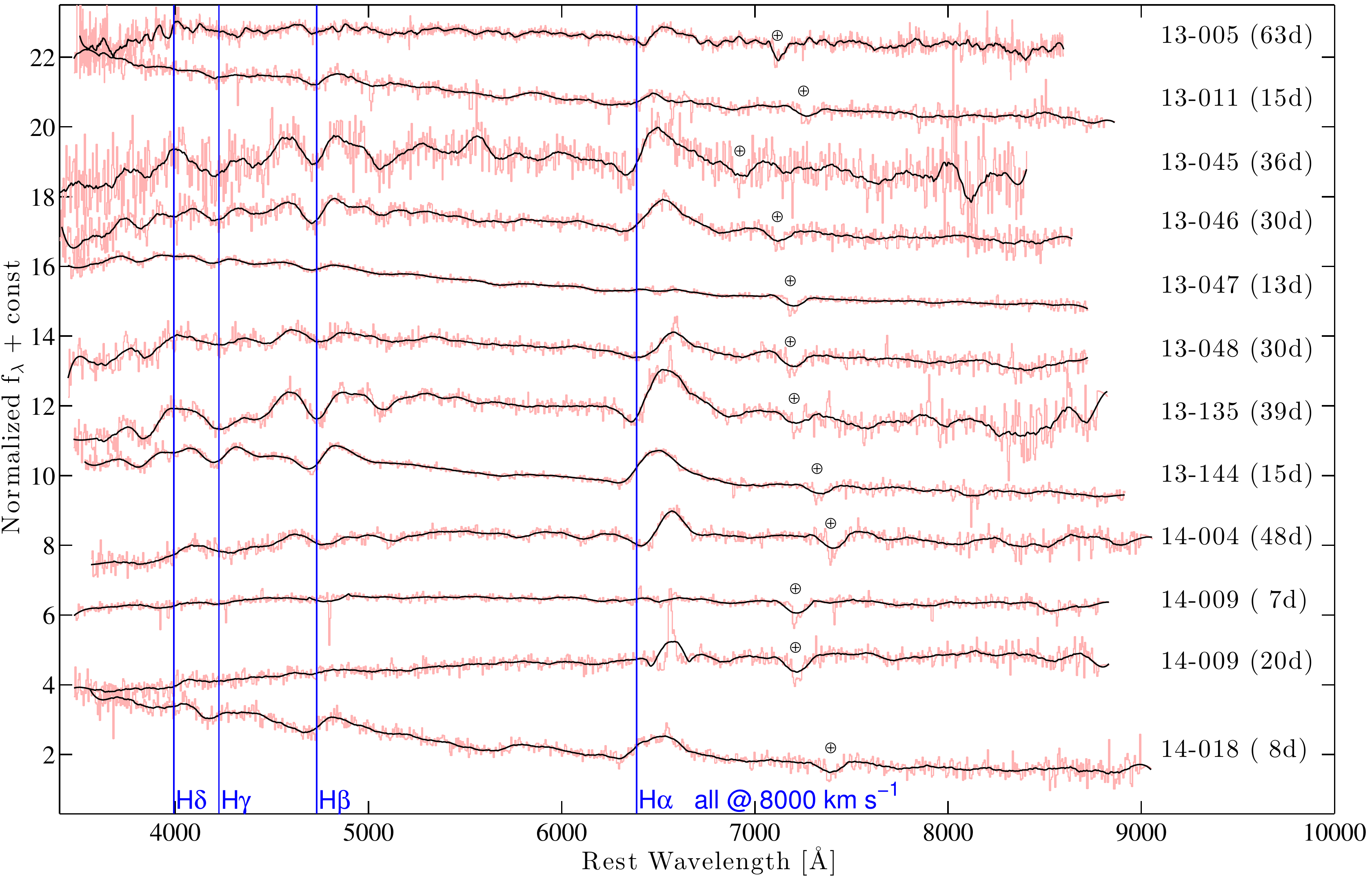}
\caption{Rest-frame spectra of the sample (pink), overlaid with smoothed curves (black), with their phase (as derived from the photometry) in parentheses. We mark the telluric feature at 7614\,\AA, and the Balmer series offset by 8000 km\,s$^{-1}$.}\label{f:spec}
\end{figure*}

\section{Introduction}\label{s:intro}

Type II supernovae (SNe II), are perhaps the simplest and in some ways best understood stellar explosions. We know they result from the core collapse of massive stars, those with masses near the 8--20\,\Msol\ range, most securely through archival progenitor detections \citep[see review by ][]{smartt09}, we know that their ejecta are composed of mostly hydrogen \citep[see review of SN spectroscopic types by][]{filippenko97}, and from extensive observations and modeling we seem to have a fair understanding of most of their photometric and spectroscopic evolution. 

Nevertheless, many questions remain unsatisfactorily answered, from the mechanism leading to their successful explosion which is still largely mysterious (e.g., \citealt{bruenn14}, and references therein), to details regarding their light curve shapes, and distribution of shapes, through the fate of the more massive of these stars, near 20\,\Msol \citep{smartt09}. 

With advances in detector technology and computing, SNe, once a  scarce commodity, are now observable in large numbers, and the field is evolving from detailed discussions of single objects, to samples that are analyzed in bulk. Several such samples of SNe II have been recently analyzed via various means \citep[e.g.,][]{arcavi12,maguire12,faran14,anderson14,faran14b,sanders14,spiro14}. However, with the availability of greater datasets some questions have actually become muddier. For example, \citet{arcavi12} find a clear gap in the distribution of light-curve decline rates, between standard plateau-like SNe II-P and declining SNe II-L, as well as rather uniform plateau durations for SNe II-P (near 100\,d). In contrast, \citet{anderson14} who use a bluer photometric band, find a continuum of decline rates, removing the ability to separate the SNe II-P from the II-L (see similar result by \citealt{sanders14}). \citet{faran14} and \citet{faran14b} find uniform plateaus but show that the gap in declines emerges largely from the analysis method. \citet{sanders14} find rather uniform durations, if somewhat shorter and with outliers ($90\pm10$\,d). While some of the differences are semantic, and others at least partly arise from methodological differences (different bands, or definition of the plateau duration), the samples do often seem different, which is puzzling. 

In this short paper we attempt to address some of these discrepancies using yet another independent sample. The SNe presented here have all been detected by the Transient Detection System of the OGLE-IV survey \citep{kozlowski13}, and are a subset of the SNe presented in \citet{wyrzykowski14}. Briefly, in OGLE-IV about 650 deg$^2$ are observed with an average 5\,d cadence. An automated pipeline finds transients down to $\sim 20\,$mag using image subtraction (see more details in \citealt{wyrzykowski14}). Using 11 SNe II, we examine their basic observational parameters -- such as light curve shapes, ejecta velocities, luminosities -- in the context of other recently published samples -- and attempt to reconcile often-conflicting findings.

\begin{table}
	\tiny
	\tabcolsep=0.11cm
\caption{SN Sample}\label{t:sample}
\begin{tabular}{lcccc}	
\hline \hline
SN name & $z_{\mathrm{host}}$ & $\mu$(mag)\tablenotemark{a} & E(B-V)$_\mathrm{MW}$\tablenotemark{b} & Explosion (MJD) \\
\hline
OGLE13-005 & 0.07 & 37.47 & 0.028 & 2456241.7 $\pm$ 2.9 \\
OGLE13-011 & 0.05 & 36.71 & 0.047 & 2456298.25 $\pm$ 6.45 \\
OGLE13-045 & 0.10 & 38.29 & 0.030 & 2456483.4 $\pm$ 2.5 \\
OGLE13-046 & 0.07 & 37.47 & 0.028 & 2456489.9 $\pm$ 2.0 \\
OGLE13-047 & 0.06 & 37.12 & 0.029 & 2456505.3 $\pm$ 4.5 \\
OGLE13-048 & 0.06 & 37.12 & 0.038 & 2456489.4 $\pm$ 1.5 \\
OGLE13-135 & 0.057 & 37.00 & 0.157 & 2456620.65 $\pm$ 2.05 \\
OGLE13-144 & 0.04 & 36.21 & 0.113 & 2456635.7 $\pm$ 6.0 \\
OGLE14-004 & 0.03 & 35.56 & 0.067 & 2456660.3 $\pm$ 2.5 \\
OGLE14-009 & 0.056 & 36.96 & 0.068 & 2456688.2 $\pm$ 1.5 \\
OGLE14-018 & 0.03 & 35.56 & 0.087 & 2456702.2 $\pm$ 1.5 \\
\hline
\end{tabular}
\tablenotetext{a}{Distance modulus assuming concordance cosmology \citep{planck-collaboration14}.}
\tablenotetext{b}{Milky Way extinction from \citet{schlafly11}.}

\end{table}

\section{Spectroscopy and Sample selection}\label{s:spec}

Out of the SNe in \citet{wyrzykowski14}, we focus on objects that can be spectroscopically classified as Type II, either II-P or II-L, avoiding SNe of Type IIn and IIb. We further excluded objects with a light curve that was clearly IIb-like (i.e., similar to SN\,1993J, \citealt{richmond94}). Since most of our objects only have one spectrum, and it was often taken early, we may have some interloping SN IIb in the sample. Most of the spectra were acquired by the PESSTO project \citep{smartt14}, and downloaded from WISEREP \citep{yaron12}, except for OGLE13-005\footnote{For compactness, we somewhat shorten the names. OGLE-SN-20XX-YYY is shortened to OGLEXX-YYY.} which was observed and classified by \citet{prieto13}. The SN list can be seen in Table \ref{t:sample}, and the spectra in Figure \ref{f:spec}, with the phases as determined from the photometry (see below). On top of the often-noisy spectra we overplot the result of smoothing them with a Savitzky-Golay filter \citep{savitzky64}. Redshifts were compiled by \citet{wyrzykowski14}, based on the spectroscopy. 

Qualitative inspection of the spectra reveals that 
OGLE-13-011, OGLE13-045, OGLE13-046, OGLE13-048, OGLE13-135, OGLE13-144, OGLE14-004, and OGLE14-018, all have broad hydrogen lines, and look like typical SNe II at their respective photometric phases. A classification with the SN Identification Program (SNID; \citealt{blondin07}) with the default templates and parameters finds the same. OGLE13-144 has the weakest ratio of H$\mathrm{\alpha}$ absorption to emission, as often seen in SNe II-L \citep{schlegel96,gutierrez14,faran14b}. OGLE14-009 is featureless. The narrow hydrogen emission in one of the spectra is consistent with being from the host galaxy (full width at half maximum of about 1000 km\,s$^{-1}$). There are no obvious spectral indications it is securely a type II, though SNe II-L sometimes develop lines only later in their evolution \citep[e.g., ][]{faran14b}. We tentatively keep it in the sample for further discussion, with the possibility it is not a type II-P or II-L. Due to its lack of features (that is intrinsic, and not due to a signal-to-noise issue, the continuum is clearly detected) SNID is unhelpful in this case. OGLE13-047 is best fit by SNID to the historical SNe II-L, SNe 1979C and 1980K, has a strong H$\mathrm{\beta}$ absorption, but no H$\mathrm{\alpha}$. This is reminiscent of the early spectra of the SNe II-L, 2001fa and 2005dq, and the superluminous II-L SN\,2008es who developed lines late in their evolution, starting from H$\mathrm{\beta}$ as well \citep{faran14b}. 

Based on spectral properties, we therefore find that the sample is indeed composed of SNe II, as constructed, most of them `regular' SNe II-P, while OGLE13-047,  OGLE13-144, and perhaps OGLE14-009 are the most II-L like. As we show below, these are also the most declining objects. 

We measure the ejecta velocities of all the SNe (except for OGLE14-009 which has no features to fit), and use them in section \ref{s:corr}. Traditionally, the Fe\,II $\lambda5169$ absorption line velocity, as measured in mid-plateau, is considered a good proxy for the velocity of the photosphere (e.g., \citealt{schmutz90}; \citealt{dessart05a}). Since the spectra have typically low signal-to-noise ratios (S/N), and were often taken early, when the line has not yet developed much, we measure the Fe\,II velocity indirectly. Using the same method as \citet{poznanski09}, \citet{poznanski10} and \citet{poznanski13}, we cross correlate the spectra, focusing on the area bluer of H$\alpha$ -- dominated in early spectra by H$\beta$, and later spectra by the Fe\,II line -- with a library of high S/N spectra for which the velocity of the $\lambda5169$ line has been measured directly. As shown by \citet{poznanski10} and \citet{faran14}, the H$\beta$ velocity is linearly related to the Fe\,II velocity. The velocity from the cross-correlation and its uncertainty are then propagated to day 50 past explosion, following \citet{nugent06}, who showed that photospheric velocities of SNe II-P follow a tight power law relation. 
We use the improved determination of the phase dependance of the velocity by \citet{faran14}. We note that \citet{faran14b} found that SNe II-L have a different, slower, velocity evolution, with some scatter. As a result, our calculations probably underestimate the velocity for such SNe. 

\begin{figure*}
\centering
\includegraphics[width=1\textwidth]{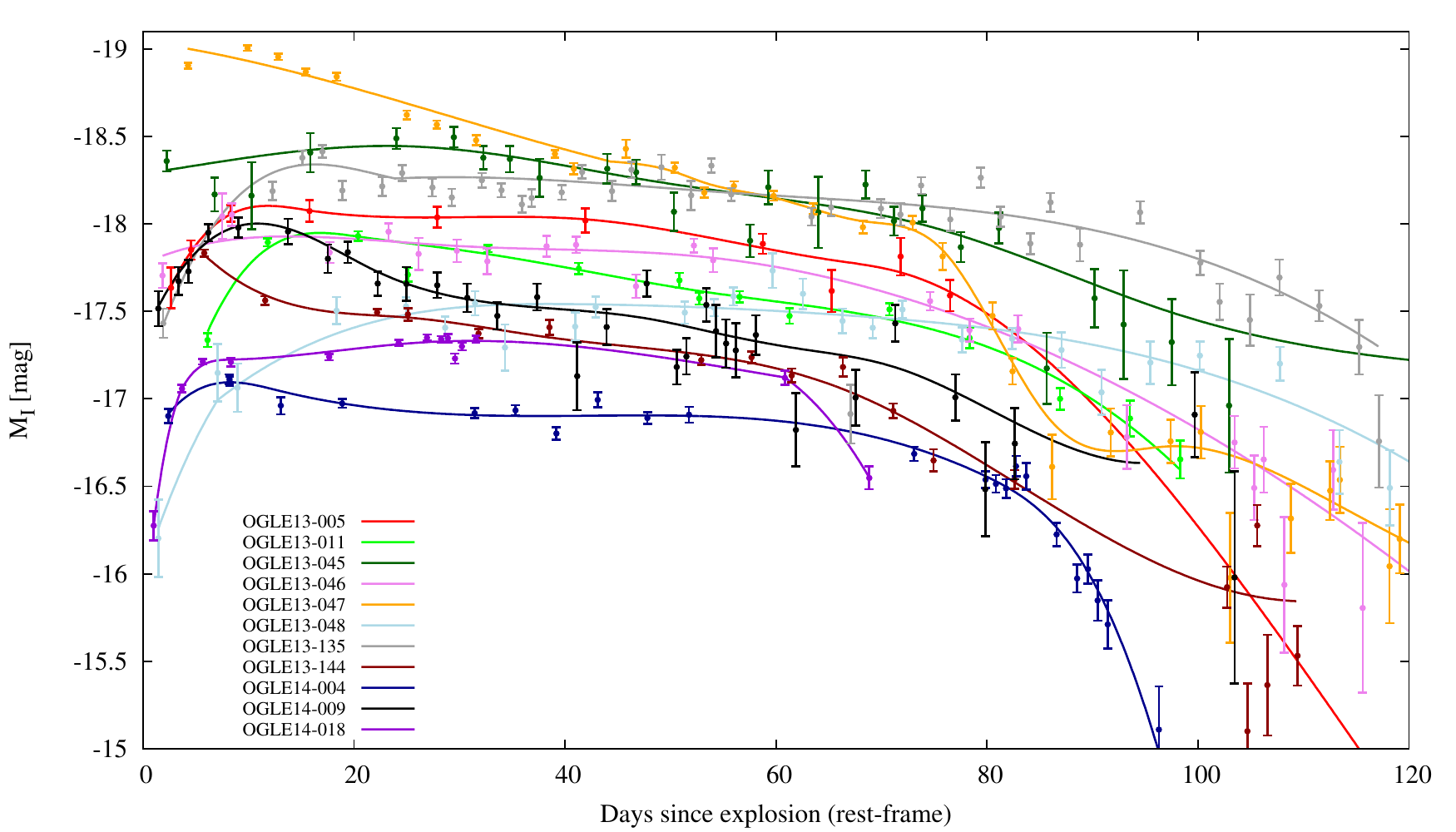}
\caption{Absolute magnitude light curves of the OGLE sample, as well as our best fitting spline fits. We crop the figure at 120\,d, the period most relevant to this study, though some SNe have light curves that extend beyond these limits. We use the complete light curves for the spline fitting.}\label{f:mag_range}
\end{figure*}

\section{Photometric Properties}

In Figure \ref{f:mag_range} we show the rest-frame $I$-band light curves of the sample (Table \ref{t:phot} includes the photometry). The photometry derived by \citet{wyrzykowski14} is corrected for galactic extinction using the maps of \citet{schlafly11}, is offset in time to match at the SN explosions days, and K-corrected using the spectra of SN\,1999em to determine the correction at every phase. The explosion days were determined as the mid point between the first detection and the last non-detection. Since the target cadence of OGLE is 5\,d, the typical uncertainty is 2\,d. We also fit a spline curve to each SN. One can see in Figure \ref{f:mag_range} that the splines capture well the variability timescales of the various light curves. 

While the SNe have a broad range of luminosities, spanning about two magnitudes, the sample is more narrowly distributed and the SNe are brighter than samples found in nearby galaxies, such as the samples recently discussed by \citet{arcavi12}, \citet{anderson14}, or \citet{faran14}. OGLE12-047, which has a single early spectrum similar to the superluminous SN\,2008es as discussed above, is also the brightest object in our sample, with a peak magnitude of about M$\sim-19$\,mag, indicating that these two SNe are somewhat similar. 

\begin{figure}
\centering
\includegraphics[width=1\columnwidth]{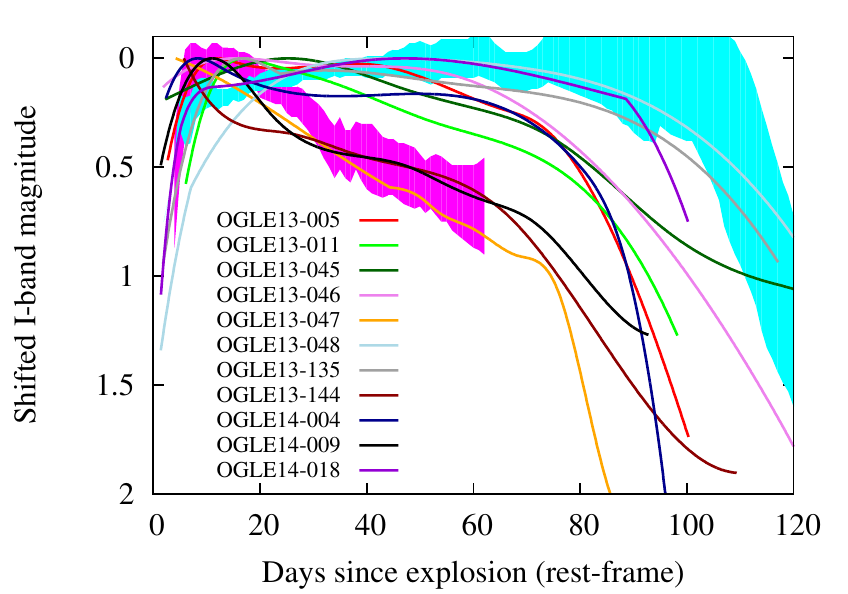}
\caption{Light curves normalized at peak luminosity compared to the $I$-band templates of \citet{faran14b}. We find 3 SNe that decline about 0.5\,mag in 50\,d, while the rest of the sample has a marked plateau, albeit a short one.}\label{f:mag_norm}
\end{figure}

If we also normalize the SNe to have the same peak magnitude, using the splines to determine the peak (a slight but essential modification to the recipe of \citealt{arcavi12} where SNe II-P and II-L were treated differently a-priori), as seen in Figure \ref{f:mag_norm}, the range of decline rates become apparent, and a minor gap may be seen to emerge between the more or less declining SNe. Clearly, the three declining SNe, OGLE13-047, OGLE13-144, and OGLE14-009, are also the most II-L like spectroscopically, as discussed above. Furthermore, OGLE-13-011, which somewhat fills that gap, has the largest uncertainty on its explosion date. Shifting it back by $\sim 6$ days would make it consistent with the SNe II-L in our sample and with the template from \citet{faran14b}, thus clearing the gap further.

This gap is reminiscent of the results found by \citet{arcavi12}, and contrasts with the findings of \citet{anderson14}, \citet{faran14b}, and \citet{sanders14}. The existence of such a gap, or its absence, could indicate whether there is a continuum in the properties of SNe II progenitors. However, often these different works are difficult to compare because their samples were obtained in different bands (e.g., $V$-band for \citealt{anderson14}, $R$-band for \citealt{arcavi12}). When we compare the OGLE light curves to the $I$-band templates of \citet{faran14b}, it appears that the samples are different. While the II-L SNe in our sample match the template reasonably well, the SNe II-P have markedly shorter-duration plateau. We measure the plateau durations following the definition from \citet{faran14}, from the date of explosion to the phase at which there is a $0.5$\,mag decline from average plateau magnitude (the average is calculated between days 25 and 75).

Most SNe II-P, including most of our sample, reach their peak luminosity quickly, within a week or so from explosion (see \citealt{faran14} for a recent compilation of rise times, as well as \citealt{valenti14} for a recent slow riser). However, OGLE13-048 reaches peak brightness late in its evolution, about 20\,d from explosion. This is reminiscent of the rare explosions of blue supergiants, such as the canonical SN\,1987A or SN\,2000cb \citep{kleiser11}, but SN\,1987A reached peak $I$-band magnitude about 80\,d after explosion, and SN\,2000cb took roughly 60\,d. In 87A-like explosions the lack of luminosity early on is attributed to the small radius of the progenitor -- the energy is expended on expansion -- and the later luminosity is dominated by the decay of $^{56}$Ni. This explanation cannot be summoned here, since 20\,d past explosion is too early for the $^{56}$Ni to peak (or would require the Ni to be located mostly far in the outer ejecta, which is unreasonable). 

OGLE14-009, with its intrinsically featureless spectra, has a declining light curve. Out of caution, we compare its light curve to a SN Ia, using synthetic photometry on the Ia templates of \citet{nugent02} as in, e.g., \citet{poznanski07}. We find that while the brightness is broadly consistent, the light curve of OGLE14-009 is significantly broader than that of a typical SN Ia, requiring a stretch of about 60 percent (which then does not fit the rising part of the light curve), or a $\Delta \mathrm{M}_{I,15}\sim 0.2$. This is broader than the unusually slowly declining SN\,2001ay \citep{baron12}. It is therefore unlikely that OGLE14-009 is a SN Ia (or any other SN type with a nickel driven light curve), given its spectra and photometry. It could still possibly be a SN IIn, as these can have wildly different light curves (e.g., \citealt{miller10}), and featureless spectra at early times (e.g. SN 1998S, \citealt{fassia01}). 

\begin{table}
	\tabcolsep=0.11cm
\caption{K-Corrected Rest-Frame Photometry}\label{t:phot}
\begin{tabular}{lcccc}	
\hline \hline
SN name\tablenotemark{a} & Phase & $I$ & $\Delta_I$ \\
\hline
OGLE14-018 & 1.47270 & 19.438 & 0.083 \\
OGLE14-018 & 5.37956 & 18.653 & 0.022 \\
OGLE14-018 & 8.25974 & 18.496 & 0.016 \\
OGLE14-018 & 12.12427 & 18.496 & 0.025 \\
OGLE14-018 & 25.70538 & 18.457 & 0.018 \\
OGLE14-018 & 35.29636 & 18.371 & 0.017 \\
OGLE14-018 & 39.20328 & 18.338 & 0.018 \\
OGLE14-018 & 41.15068 & 18.345 & 0.018 \\
OGLE14-018 & 42.08987 & 18.339 & 0.024 \\
OGLE14-018 & 43.06376 & 18.455 & 0.029 \\
OGLE14-018 & 44.01478 & 18.384 & 0.022 \\
OGLE14-018 & 45.95610 & 18.341 & 0.019 \\
OGLE14-018 & 88.60024 & 18.540 & 0.041 \\
OGLE14-018 & 100.24717 & 19.107 & 0.065 \\
OGLE14-018 & 182.25189 & 20.740 & 0.167 \\
\hline 
OGLE14-009 & 1.44567 & 19.641 & 0.100 \\
OGLE14-009 & 3.36477 & 19.482 & 0.079 \\
OGLE14-009 & 4.30878 & 19.425 & 0.065 \\
OGLE14-009 & 6.19501 & 19.202 & 0.049 \\
OGLE14-009 & 9.03278 & 19.169 & 0.057 \\
OGLE14-009 & 13.74580 & 19.185 & 0.073 \\
OGLE14-009 & 17.52362 & 19.335 & 0.081 \\
OGLE14-009 & 19.41935 & 19.295 & 0.061 \\
OGLE14-009 & 22.25030 & 19.471 & 0.069 \\
OGLE14-009 & 24.97752 & 19.470 & 0.096 \\
...\\

\end{tabular}
\tablenotetext{a}{Full table in online version.}
\end{table}

\section{correlations}\label{s:corr}

Whether or not there is a gap between SNe II-L and SNe II-P, recent studies indicate that Type II SNe (barring SNe IIb and IIn) form a one-parameter family. Brighter SNe have higher ejecta velocities, and more declining light-curves \citep[e.g., ][]{hamuy02,poznanski09,anderson14,sanders14}. Furthermore, \citet{poznanski13} finds that the brighter SNe come from more massive progenitors (see also \citealt{smartt09}) that have had a much greater energy deposited by the explosion in the envelope, so that the energy $\mathrm{E}$ scales as $\mathrm{M}^3$, where $\mathrm{M}$ is the initial mass of the progenitor. Combining these findings with the numerical simulations of \citet{dessart10a}, it appears that brighter SNe should have shorter plateaus, down to about 80\,d for stars above $20$\,\Msol. Therefore mass determines the energy, luminosity, velocity, plateau duration, and decline rate. Note however that \citet{poznanski09} find that declining SNe do not obey the luminosity-velocity relation found by \citet{hamuy02}, (and as mentioned above, their velocity evolution might also differ) which may indicate that they do not follow other scalings either. 

\begin{figure}
\centering
\includegraphics[width=1\columnwidth]{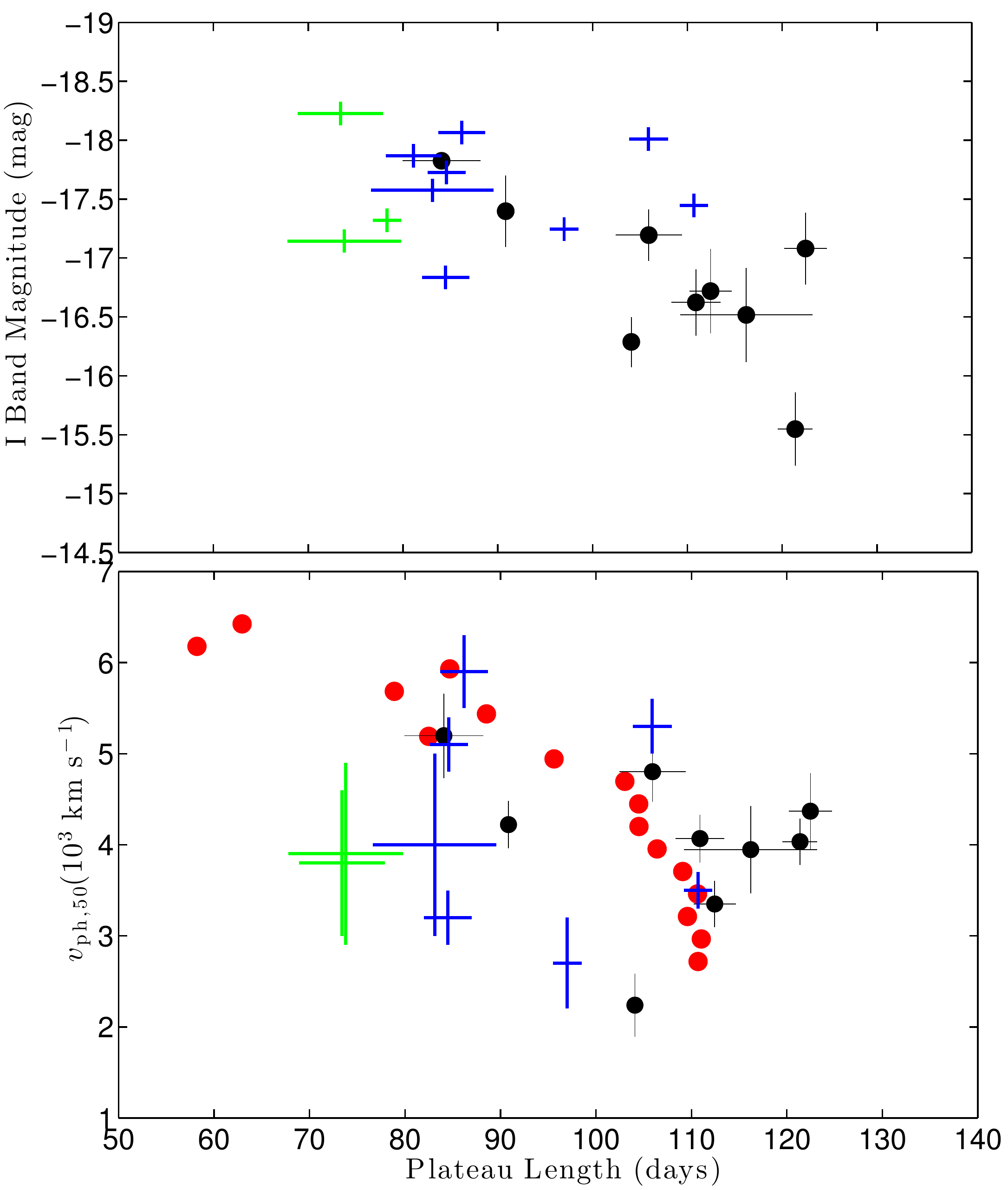}
\caption{Correlation of velocity (bottom panel) and $I$-band magnitude (top) with plateau duration. Black points from \citet{faran14}, red points from the numerical models of \citet{dessart10a} assuming  $\mathrm{E} \propto \mathrm{M}^3$\ \citep{poznanski13}, blue (green) crosses are the SNe II-P (II-L) from OGLE. While there is a weak correlation between brightness and plateau duration, the same does not apply to the velocity.} \label{f:corrs}
\end{figure}

In the top panel of Figure \ref{f:corrs} that includes the sample of SNe II-P from \citet{faran14}, as well as our OGLE sample, one can see that indeed there is a weak correlation between plateau duration and magnitude, such that only brighter events can have short plateaus, but there does not seem to be a population of faint SNe with short plateaus. However, in the lower panel one can see that a significant fraction on the OGLE SNe have low velocity, and short plateaus, contrary to the expectations from the one dimensional picture above. 

\begin{figure}
\centering
\includegraphics[width=1\columnwidth]{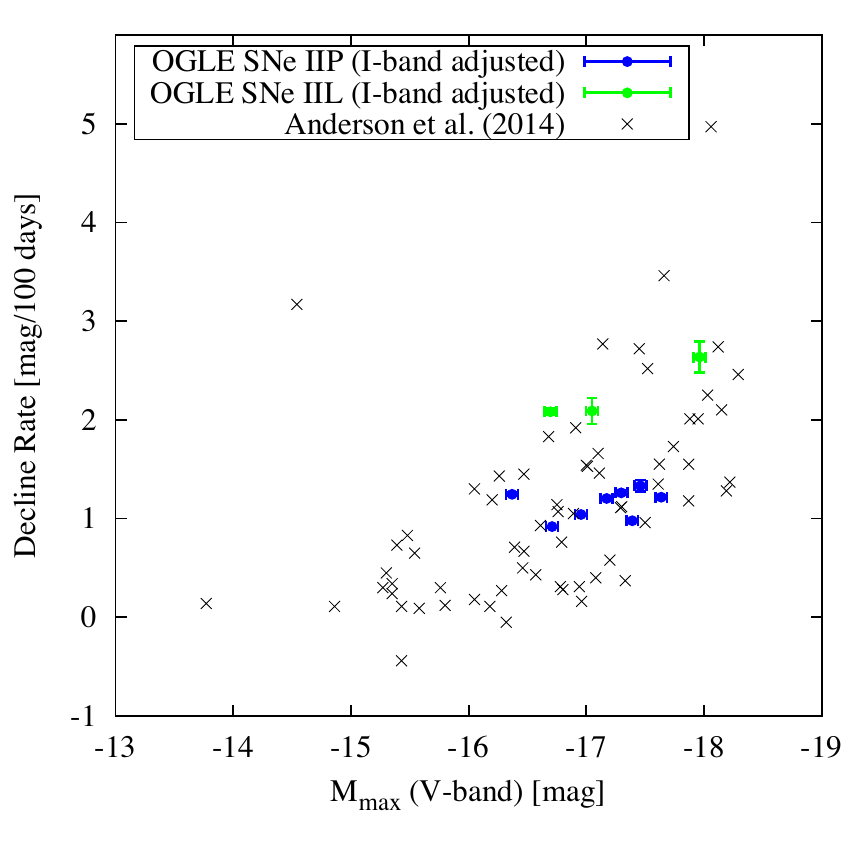}
\caption{Correlation of peak magnitude with decline rate. Grey points from \citet{anderson14}, blue (green) crosses are the SNe II-P (II-L) from OGLE. In order to compare our data to the $V$-band magnitudes in \citet{anderson14}, we multiply our $I$-band decline rates by two, while offsetting the peak magnitude by 0.5\,mag, based on the templates of \citet{faran14b}.} \label{f:dec}
\end{figure}

In figure \ref{f:dec} we examine the correlation between peak magnitude and decline rate, comparing to the findings of \citet{anderson14}. There are two difficulties when comparing these samples. First, since our light curves are not very well sampled we cannot differentiate between various phases these authors define. Instead we find the decline rate by asking at what phase $t_d$ the spline curve that was fit to every object crosses 0.5\,mag. The decline is then $50/t_d$ in units of $\textrm{mag} / 100 \textrm{d}$, where the uncertain explosion date dominates the uncertainty. Secondly, our light curves are in $I$-band, while \citet{anderson14} only studied $V$-band data. Using the templates from \citet{faran14b}, we find that SNe II have a peak color of $V-I = 0.4-0.7$, and SNe II-L decline about twice as faster in $V$ than in $I$. For this qualitative comparison we therefore apply an offset of 0.5\,mag to our peak brightness, and scale our declines by a factor of 2. Our objects seem in agreement with the sample of \citet{anderson14}.

\begin{figure*}
\centering
\includegraphics[width=.9\textwidth]{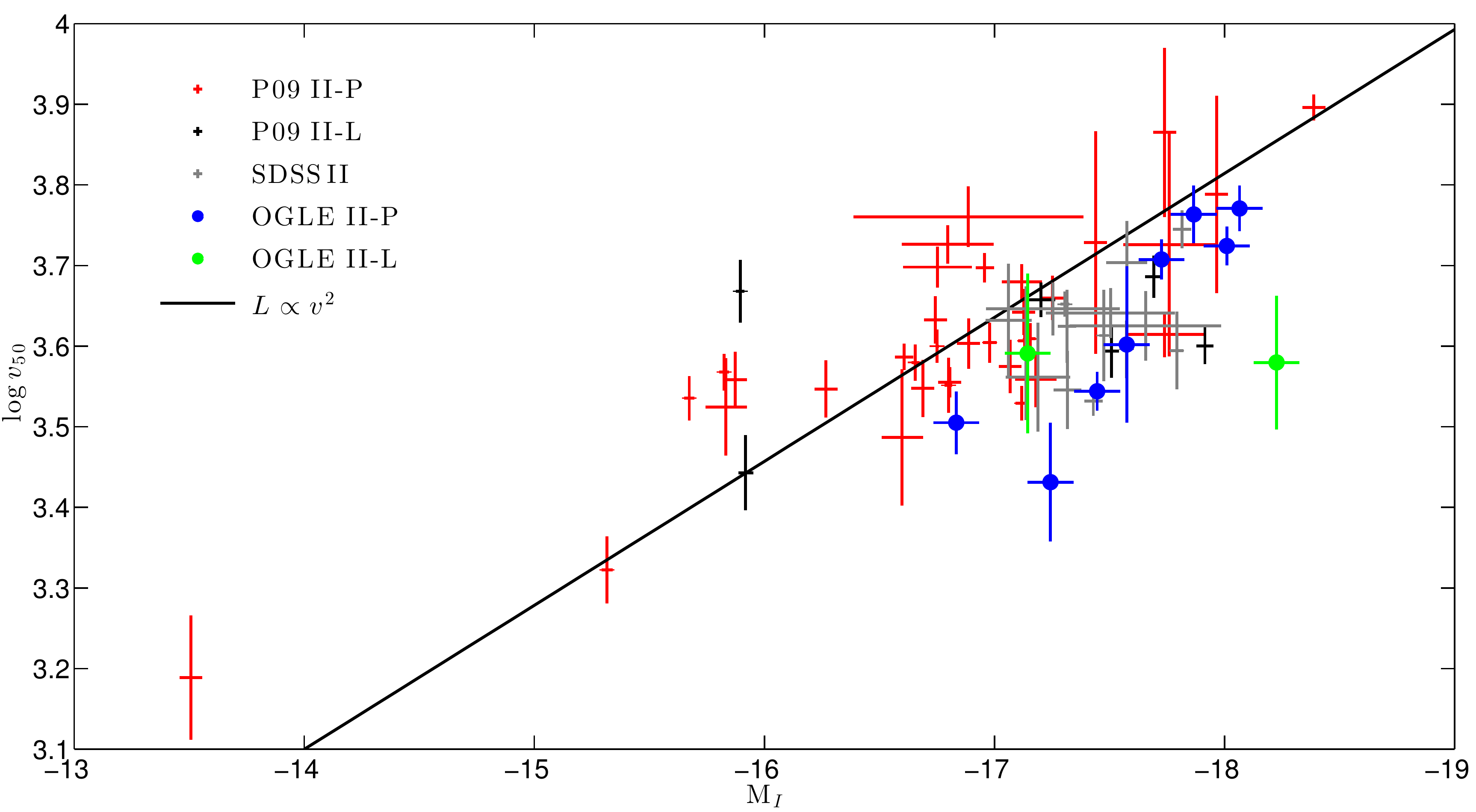}
\caption{Luminosity-velocity relation. Samples from \citet{poznanski09} in red (SNe II-L in black), from SDSS\,II \citep{poznanski10} in grey, and OGLE in blue (SNe II-L in green). The SNe from magnitude limited surveys (SDSS\,II and OGLE) are clearly offset, though they may follow their own luminosity-velocity relation.} \label{f:corrs2}
\end{figure*}

In figure \ref{f:corrs2} we study the `Hamuy \& Pinto' velocity-luminosity relation. We compare our sample to the sample compiled by \citet{poznanski09} -- which includes data from \citet{hamuy02} and \citet{nugent06} as well as a subset of the SNe in \citet{faran14} and \citet{faran14b}\footnote{SN\,2002hh was omitted due to its abnormally high extinction.}, and the SDSS\,II sample from \citet{poznanski10}, which is a reanalysis of the sample of \citet{dandrea10}. The line shows the best fit luminosity-velocity relation, as derived by \citet{poznanski09}. Clearly the OGLE SNe are all over-luminous, or have slow ejecta, when compared to most of these samples, as all of the objects lie beneath the line, similarly to the SDSS\,II sample. Also, it seems that the decline rate is a weak indicator of fit quality as declining objects (from either samples) do not seem particularly scattered or biased, though if taken individually, they do not seem to follow the relation at all. One should bear in mind that this plot does not take into account any color information -- a tracer of dust extinction and intrinsic variance -- that is typically used to reduce the scatter in such diagrams. All luminosities here are under the assumption of no significant extinction in the host galaxy. 

Surprisingly though, excluding the declining objects, the OGLE and SDSS\,II SNe do seem to have a luminosity-velocity correlation, albeit it is offset from the one derived from nearby samples. While this could be a dust bias -- nearby samples have more dusty SNe which appear fainter on this diagram --  as was shown to be possible for SDSS\,II sample \citep{poznanski10}, we do not have color information for the OGLE sample to test this hypothesis fully. 
We do however search for Na\,I\,D absorption in all of the spectra and find none (though this is only a weak indicator of extinction, as shown by \citealt{poznanski11}, and our S/N is typically low). As mentioned before, the velocity derivation, using the power-law behavior found by \citet{nugent06}, was shown not to be applicable to SNe II-L by \citet{faran14b}. These SNe appear to evolve more slowly, so that their velocity here might be underestimated. Correcting for this possible bias would somewhat increase their velocity.

\section{Discussion}

Examining the light curve of OGLE13-047, our brightest and best observed SN II-L, one can see that after a period of decline of about $80$\,d, it goes through a second, sharper, drop, akin to the falling-off the photospheric phase of SNe II-P. Since it it also brighter than typical SNe II-P, it could be perhaps explained with a similar model but with a additional energy source that supplies an extra hump on top of the plateau. Alternatively, a different profile of the stellar envelope before the explosion could perhaps account for it. Recently \citet{goldfriend14} showed that the shape of a Type II light curve depends strongly on the mass profile, with the plateau resulting from a somewhat fortuitous coincidence with hydrogen recombination of a narrow range of profiles. A late drop akin to the falling of the plateau reinforces such a picture where the early light curve is dominated by the bulk of the envelope: its profile, and its composition. See similar recent observations by \citet{valenti15}.

OGLE13-047 reaches $-19$\,mag, about 40 times brighter than under-luminous SNe II-P such as SN\,2005cs. Assuming the physics driving the light-curves of SNe II-L and II-P are similar, This large span is hard to reconcile with the analytical findings of \citet{popov93}, or more recently \citet{goldfriend14}, where the luminosity during the photospheric phase is found to be only weakly dependent on the various parameters. The range observed would require a deposited energy 100 times larger in luminous events, radii 1000 times larger, wildly different opacities, or any combination thereof. However, if the energy strongly depends on the progenitor mass, as found by \citet{poznanski13}, one obtains a luminosity that depends quadratically on the mass, alleviating some of the difficulty to explain this range of luminosities, though not all. 

We find 2 or 3 SNe II-L, out of a sample of 11. A II-L fraction of $\sim 30$ percent could be surprising considering their scarcity in nearby searches \citep{li11,faran14b}, but the difference between the OGLE sample and nearby samples can be explained as stemming from the different selection biases influencing different SN searches. Nearby searches are more complete in luminosity, sensitive to much fainter SNe. The effective survey volume for SNe near $-15$\,mag (like SN\,2005cs for example), is about 250 times smaller than for SNe reaching $-19$\,mag. Magnitude limited surveys are severely Malmquist biased, finding the brightest objects of a given distribution, even when intrinsically rare. Furthermore, nearby searches are typically focused on massive, luminous, star-forming, spiral galaxies, preferring SNe that would occur in such metal rich environments. This could also change the ratio of SN types \citep{arcavi10}.

The OGLE sample, like the SDSS\,II sample from \citet{dandrea10} before it, is therefore biased towards brighter SNe that have shorter plateaus (or an actual decline), but their velocities do not match the expectations from nearby searches. The sample from \citet{sanders14} seems similarly biased. In their Figure 14 one can see that their SNe peak brighter, around --18\,mag, depending on the band, and decline rapidly, as seen in their Figure 11. In fact, the majority of their SNe would be SNe II-L by the \citet{li11} criterion of 0.5 mag\,/\,50\,d in R. The brightest SNe in the sample from \citet{anderson14} also have short plateaus, driving their mean closer to ours.

\section{Conclusions}

Having analyzed a sample of 11 SNe from the OGLE-IV survey, we find that two or three of them are SNe II-L. They are distinct both in their decline and in their spectroscopic properties as previously suggested. 

The 8 SNe II-P while rather standard in most respects, are more luminous than typically found by nearby searches, have shorter-duration plateaus, and rise times that can be as long as 20\,d, more than double the typical timescale.

We therefore find that a single parameter cannot explain the range of outcomes from massive hydrogen-rich core collapse events. While the diversity for a subset of these can be shown to follow mass, searches that are more sensitive to bright SNe find objects that behave differently. A complete census of core collapse should therefore rely on a careful combination of samples from both magnitude and volume limited surveys, or a very wide un-targeted search, such as the Palomar Transient Factory (PTF; \citealt{law09}) that can also find a substantial number of faint events and account for them properly.

\section*{Acknowledgments}
We thank the PESSTO collaboration and J. Prieto for the spectra used in this paper, and  J. Anderson, I. Arcavi, and N. Sanders for comments on this manuscript. We further thank the referee for their review of this work. 

D.P. acknowledges support from the Alon fellowship for outstanding young researchers, and the Raymond and Beverly Sackler Chair for young scientists. This work was partially supported by the Polish Ministry of Science and Higher Education through the program ``Ideas Plus'' award No. IdP2012 000162.

This research made use of the Weizmann interactive supernova data repository  (\texttt{www.weizmann.ac.il/astrophysics/wiserep}), as well as the NASA/IPAC Extragalactic Database (NED) which is operated by the Jet Propulsion Laboratory, California Institute of Technology, under contract with NASA. 
This work is based on observations collected at the European Organisation
for Astronomical Research in the Southern hemisphere,
Chile as part of PESSTO (the Public ESO Spectroscopic Survey for
Transient Objects Survey) ESO programme ID 188.D-3003.
The research leading to these results has received funding from
the European Union Seventh Framework Programme (FP7/2007-
2013) under grant agreement no. 264895.

\bibliography{OGLEbib}
\bibliographystyle{mn2e} 
\end{document}